# Atomic-scale investigation of hydrogen distribution in a Ti-Mo alloy

Fengkai Yan[1,2], Isabelle Mouton[1], Leigh T. Stephenson[1], Andrew J. Breen[1], Yanhong Chang[1], Dirk Ponge[1], Dierk Raabe[1], Baptiste Gault[1]

*1. Max-Planck-Institut für Eisenforschung GmbH, Max-Planck-Straße 1, 40237, Düsseldorf, Germany*

*2. National Laboratory for Materials Science, Institute of Metal Research, Chinese Academy of Sciences, Shenyang 110016, People's Republic of China*

*Corresponding authors: fkyan@imr.ac.cn; b.gault@mpie.de*

**Abstract**

Ingress of hydrogen is often linked to catastrophic failure of Ti-alloys. Here, we quantify the hydrogen distribution in fully β and α+β Ti-Mo alloys by using atom probe tomography. Hydrogen does not segregate at grain boundaries in the fully β sample but segregates at some α/β phase boundaries with a composition exceeding 20 at.% in the α+β sample. No stable hydrides were observed in either sample. The hydrogen concentration in β phases linearly decreases from ~13 at. % to ~4 at. % with increasing Mo-content, which is ascribed to the suppression of hydrogen uptake by Mo addition.

The β-titanium alloys can absorb large amounts of hydrogen, even up to 50 at. %, which unavoidably leads to environmental hydrogen-induced cracking, i.e. hydrogen embrittlement [1-4]. In order to improve the yield strength of such alloys, some fine α-phases are often introduced by ageing. These aged alloys are more prone to hydrogen embrittlement which was ascribed to the presence of phase boundaries and α-precipitates [4-7]. Early investigations [4-6] speculated that hydrogen-assisted cracking was involved because of hydrogen trapping at β grain boundaries or at α/β interfaces, the formation of brittle hydrides at boundaries, or the hydrogen partitioning between the α and β phases. However, elucidating what mechanisms control hydrogen embrittlement in β-titanium alloys is still an open problem, because characterizing the hydrogen distribution associated with the previously mentioned microstructural features is still challenging at high compositional and spatial resolutions. Directly observing hydrogen within these features is difficult due to its high diffusivity and light weight [1]. Several conventional hydrogen detection methods, such as thermal desorption spectroscopy [8], secondary ion mass spectrometry [8, 9] and scanning kelvin probe force microscopy [8, 10], have been proposed for study of hydrogen trapping, but these methods often provide limited micrometer-scale bulk-averaged microstructural information.

Recent studies [11-15] reported that atom probe tomography (APT) is a very promising technique to provide three-dimensional quantification and visualization of the hydrogen distribution with near-atomic resolution. Through isotopic marking (i.e. using deuterium) and cryogenic transfer, Chen et al. [11] and Takahashi et al. [12] directly observed three-dimensional mapping of deuterium in vanadium carbide strengthened steels. Chang et al. [15] also investigated the hydrogen distribution in pure-Ti and Ti-alloys by APT and demonstrated the presence of titanium hydride along both α grain boundaries and α/β phase boundaries. Insights from APT pave the way to a better understanding of hydrogen embrittlement. Here, we use APT to investigate the distribution of hydrogen in a binary metastable β Ti-6 at. %Mo (Ti-12 wt. %Mo) alloy that deforms through both, dislocation slip and combined twinning-induced plasticity and transformation-induced plasticity [16]. The composition of hydrogen and molybdenum are given in atomic per cent throughout this paper unless stated otherwise.

The Ti–6Mo alloy was produced by a self-consumable melting technique. The ingot bar was hot–forged and homogenized at 1173 K for 30 min and then water quenched to obtain a fully β-phase sample (as-homogenized samples). In order to optimize the material's mechanical properties, samples were then cold rolled to a thickness reduction ratio of 40% and aged at 500 °C to obtain a dual-phase α+β phase Ti-Mo alloy (as-aged samples). Microstructures were characterized using scanning electron microscopy (SEM) on a Zeiss-FIB microscope and transmission electron microscopy (TEM) on a FEI Talos microscope. APT specimens were prepared by site specific lift-outs using a dual beam focused-ion beam (FIB) FEI Helios Nanolab 600/600i according to the procedures described in ref. [17]. After sharpening the needle-shape specimen, a final cleaning at 5 kV was performed to remove the surface beam-damaged regions. APT analyses were performed on a Cameca LEAP 3000 HR operating at a base temperature of 70 K in voltage pulsing mode with 15% pulse fraction and 200 kHz pulse frequency. Samples were not subject to a specific hydrogen charging treatment, since the Ti-alloys can absorb enough hydrogen during specimen preparation.

The microstructure of the as-homogenized Ti-6Mo alloy is fully composed of equiaxed β with a grain size of ~100 μm according to electron back-scattered diffraction mapping (EBSD) (Fig.1a). Fig. 1b-c shows an APT reconstruction from the analysis of a specimen containing a grain boundary which is decorated by carbon impurities along with the corresponding composition profile. As hydrogen atoms occupy interstitial sites in the lattice, the Ti and Mo contents were calculated by excluding the H (as shown by the left Y-axis of Ti/Mo composition in Fig. 1c) while the hydrogen composition was calculated by including the Ti and Mo (as shown by the right Y-axis of H composition in Fig. 1c), respectively. The H, Ti, Mo atoms are distributed homogeneously and the hydrogen concentration is ~26% at the grain boundary, nearly identical to the grain interior (~28%) as seen in Fig. 1c. This contrasts with α-titanium alloys in which hydrogen segregates at grain boundaries, often forming hydrides [15]. However, within some local regions of the grain interior in another specimen, the hydrogen distribution is not homogeneous (Fig. 1d-f). Hydrogen forms nanoscale elongated plate-shaped agglomerations with a composition of ~35 % (Fig. 1e). These agglomerations are not typical titanium hydrides in which the hydrogen concentration usually exceeds ~50% in terms of Ti-H phase diagram [2], such as δ-hydride (TiH$_{1.5-1.99}$) and ϵ/γ-hydride (TiH$_x$, x>1.99) [18].

Here, during FIB-based specimen preparation, significant amounts of hydrogen were introduced inside the specimen in an uncontrolled manner [15, 19]. Hydrogen composition measured by APT is usually ~10-30 times higher than the intrinsic hydrogen content in the Ti-alloy bulk material [15]. It has been proposed that during the FIB-sharpening in ~$10^{-6}$ mbar vacuum in the SEM-FIB chamber, the gallium ion beam can decompose hydrocarbons and water that had adsorbed at the metal surface and hence produce hydrogen [19]. The high affinity of Ti for hydrogen and the rapid diffusion of hydrogen can promote rapid uptake of atomic hydrogen into the specimens.

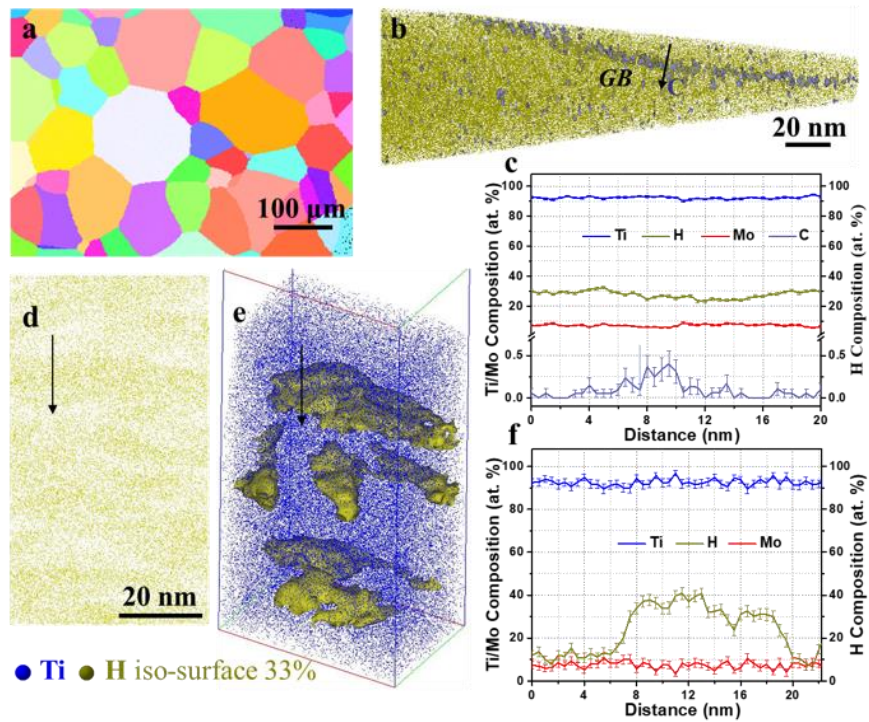

Fig. 1 (a) EBSD map showing the microstructure of the as-homogenized β Ti-6Mo alloy. (b) H (dark yellow) and C (violet) distribution map and (c) corresponding H, Ti (blue), Mo (red) and C composition profile across the grain boundary (GB, indicated by an arrow in (b)) with the decoration of C. An enlarged region of interest in another tip showing the distribution map of (d) H, (e) H using 33% H isoconcentration surface and (f) composition profile along the arrow in (d, e).

SEM observation (Fig. 2a) shows that the aged microstructure consists of two types of nanostructured regions, i.e. equiaxed grained regions and lamellar colonies. These lamellar colonies are composed of nanoscale lamellae with a high number density as revealed by scanning TEM with high-resolution high angle annular dark field (STEM-HAADF) imaging (Fig. 2b). The average lamellae spacing is ~30 nm, much smaller than the equiaxed grains (~160 nm). Energy dispersive spectroscopy (EDS) mappings (Fig. 2c-d, rectangle in Fig. 2b) reveal an inhomogeneous distribution of Ti and Mo in both nanostructured regions. The Mo-enriched grains are mainly β phase and the Mo-depleted grains are mainly α phase due to the β phase stabilizing effect of Mo [2].

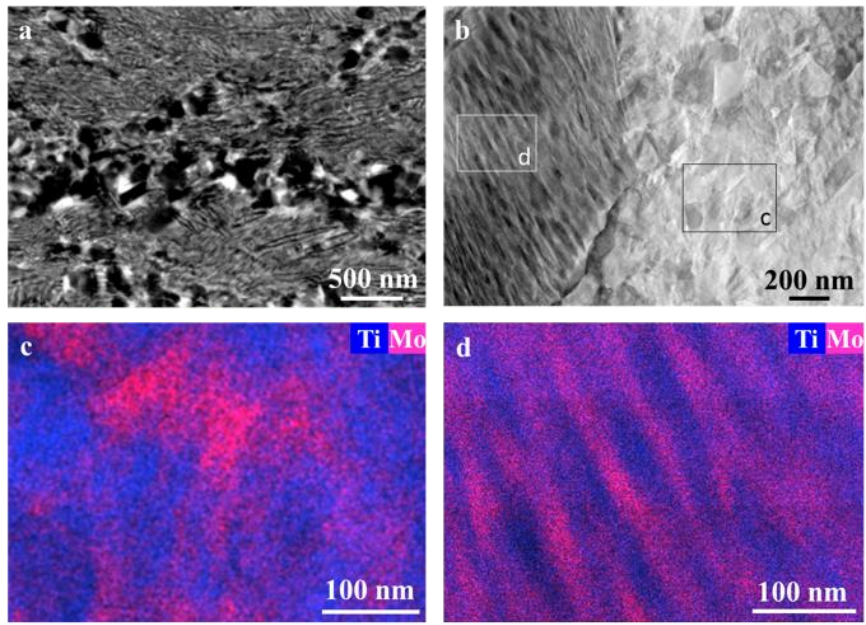

Fig. 2 (a) SEM and (b) STEM-HAADF images showing the microstructure of nanoscale equiaxed grains mixed with nanoscale lamellae colony in the α+β Ti-Mo alloy produced by cold rolling and subsequent annealing at 500 °C for 1 hour (as-aged Ti-Mo alloy). EDS maps showing the inhomogeneous distribution of Ti and Mo in (c) equiaxed grains (rectangle c in (b)) and (d) nanoscale lamellae colony (rectangle d in (b)).

Fig. 3 is an APT reconstruction showing the distribution of H, Mo, and Ti and the corresponding composition profiles in the equiaxed-grained regions. Consistent with TEM and EDS observations, Fig. 3a reveals Mo-partitioning into the β phase with concentrations ranging from 10% to 20%, whereas α is Ti-rich and only contains ~1% Mo (Fig. 3c-d). The hydrogen content in β ranges from ~6% to ~13%, i.e. much higher than the ~2% hydrogen found in α. Some hydrogen is segregated to the α/β interfaces. The composition profile across one of these interfaces, indicated by an arrow in Fig. 3a, shows that the average hydrogen composition can be as high as ~20% (Fig. 3b). The thickness of this hydrogen layer, highlighted by an H-isosurface with a threshold of 16.2%, is approx. 5 nm. These hydrogen-enriched layers are not typical titanium δ-hydride or ϵ/γ-hydride either. They also do not match the γ-hydride (TiH$_1$) compositions previously reported at α /α boundaries in Ti-2Mo [15]. Considering hydrogen being as a β phase stabilizer [2], these hydrogen-enrichment layers could be hydrogen-stabilized β phases.

The quantification of H at α, β and α/β phase boundaries measured by APT is influenced by the residual hydrogen in the APT analysis chamber, despite a low base pressure of ~ 4-6 x 10$^{-11}$ mbar. The local electrostatic field leads to the detection of various amounts of H$^+$ and H$_2^+$ ions. These ions are difficult to distinguish from the solute hydrogen within materials. Yet, variations of the residual hydrogen among different microstructural features can be roughly evaluated by estimating the change in the local electrostatic field between them [20-23]. For example, a higher concentration of residual hydrogen can be detected when the electrostatic field strength is lower during an APT voltage mode measurement [20, 21]. The field strength in turn can be estimated by calculating the relative ratio between the different charge states of e.g. Ti (i.e. Ti$^{3+}$/Ti$^{2+}$) due to the dependence of the post-ionization probability with the electrostatic surface field [24].

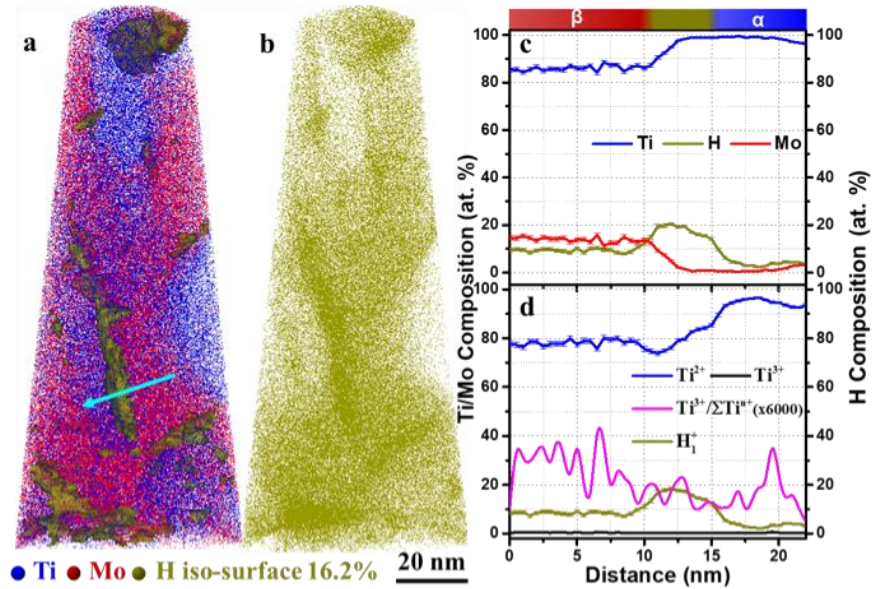

Fig. 3 Distribution maps of (a) Mo, Ti, H using 16.2% isoconcentration surface and (b) H; (c) composition profile of H, Ti and Mo, and (d) relative composition of H$^{1+}$, Ti$^{2+}$, Ti$^{3+}$ and corresponding Ti$^{3+}$/ΣTi$^{n+}$ measured through the arrow in (a) in the equiaxed α and β grained regions in the as-aged Ti-Mo alloy.

Fig. 3d shows the typical composition variation of $H^+$, $Ti^{2+}$, $Ti^{3+}$ and $Ti^{3+}/\Sigma Ti^{n+}$ across the nanoscale α and β (arrowed in Fig. 3a). Apparently, the composition of $Ti^{3+}$ and the charge-state ratio of $Ti3^+/\Sigma Ti^{n+}$ (in pink) in β is approximately two times higher than in α, evidencing a higher evaporation field of β. Therefore, the contribution of the residual H is expected to be lower in β [20, 21]. However, the hydrogen composition measured in β (~4% to ~13%) is significantly higher than in α (~2 %). It can hence be concluded that the increase in the hydrogen composition in β is related to solute hydrogen present in the material. Moreover, the composition of $Ti^{3+}/\Sigma Ti^{n+}$ at β/α interfaces does not change sharply, roughly identical to the neighboring β and α. Such phenomenon was also observed in the α+β Ti6246 alloy [15]. It indicates that the corresponding interface local field strength does not lead to significant segregation of hydrogen from the residual gas. Instead, the hydrogen segregation at interfaces mainly originate from the solute H within materials.

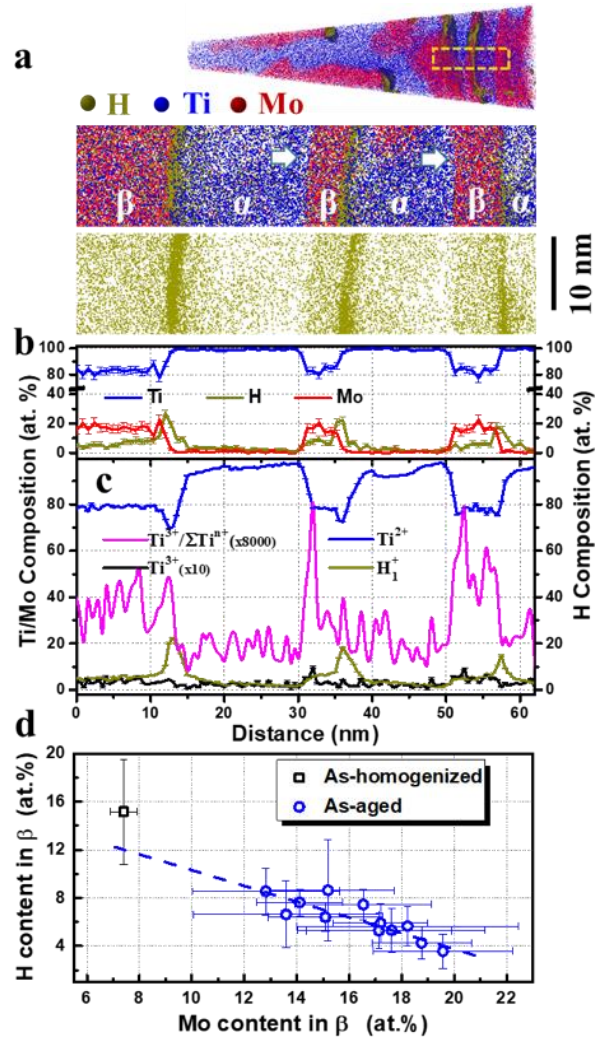

Fig. 4 An enlarged region of interest (outlined by yellow rectangle) showing the distribution of (a) Ti, Mo and H (b) the corresponding 1D composition profile across the nanoscale α and β lamellae in the as-aged Ti-Mo alloy; (c) Relative composition of $H^{1+}$, $Ti^{2+}$, $Ti^{3+}$ and corresponding $Ti^{3+}/\Sigma Ti^{n+}$ measured across the α and β lamellae. (d) Variation of hydrogen composition with the Mo content in β phases in the as-homogenized and as-aged Ti-Mo alloys.

For the nanoscale lamellae colonies, similar hydrogen partitioning effects are observed as reveled by APT analysis. As shown in Fig. 4a-b, the nanoscale lamellae are composed of multilayers of α and β phases. Within the β phase layer, the Mo concentration ranges from ~13% to ~20% and the hydrogen concentration ranges from ~4% to ~8%. In contrast, only ~0.7% Mo and ~2.4% H are detected within α phase layer. Similar hydrogen segregation is also observed at some of the β/α interfaces. Variations of hydrogen composition among the microstructural features, i.e. α, β and α/β interfaces (decorated by hydrogen segregation) in lamellae colonies, as seen in Fig. 4c, are mainly determined by the hydrogen within materials, similar with equiaxed regions as shown above Fig. 3d. However, in nanoscale lamellae colonies, the $Ti^{3+}/\Sigma Ti^{n+}$ ratio at α-to-β interfaces indicated by white arrows in Fig. 4a shows abrupt variations, as seen in Fig. 4c. This is consistent with the expected increase of the electrostatic field strength at α-to-β interfaces [20, 21]. The corresponding high electrostatic field could affect the field evaporation behavior of solute hydrogen from the material and explain that no hydrogen segregation is observed. The effects of field strength on the different segregation behavior of hydrogen at α/β interfaces needs to be further investigated.

According to the above analyses, the APT results clearly provide direct and solid experimental evidence of hydrogen partitioning in fully β and α+β titanium alloys. Firstly, the solubility of hydrogen in β phases is much higher than in α phases. This is mainly determined by the difference in crystalline structure due to that hydrogen tends to occupy tetrahedral interstitial sites [18]. The β phase shows a body-centered-cubic crystalline structure which contains 12 tetrahedral interstices while only 4 tetrahedral interstitial sites are presented in hexagonal-close packed lattice of α phase. More interstitial sites mean a higher capacity for hydrogen uptake [18]. Secondly, hydrogen segregation can occur at some α/β interfaces. Such phenomenon was also deduced in many previous investigations [1]. Kirchheim [25, 26] indicated that the chemical potential of hydrogen trapping within these defects, i.e. vacancies, dislocations and interfaces, played a central role in determining the distribution of hydrogen among various hydrogen trapping sites. Hydrogen segregation at interfaces leads to a reduction of the interface energy. Last but not least, hydrogen segregation at interfaces does in the present case not necessarily result in stable δ-, ϵ- and γ-titanium hydride formation. This is mainly due to the high hydrogen solubility of β phases and Mo alloying which suppresses the formation of hydride [15, 18].

We also find that the Mo composition significantly fluctuates in β grains due to the limited diffusion of Mo in the aged Ti-Mo alloy. Such inhomogeneity of Mo significantly influences the hydrogen uptake in β grains. As shown in Fig. 4d, the hydrogen composition roughly decreases linearly with increasing Mo-content in β grains. For the as-homogenized single β Ti-Mo alloy, the hydrogen composition is up to ~15.0% in the matrix surrounding the H-agglomerations, and overall up to 30%. In the literature, first principles calculations [27-29] suggest that due to the smaller size of Mo relative to Ti, the lattice parameter of Ti-Mo alloys decreases with increasing amount of Mo. The lattice parameter reduced from ~ 0.3264 nm for a Ti-7Mo alloy to ~0.3248 nm for a Ti-16Mo alloy [29]. Such reduction of the unit cell dimension with increasing Mo content in β also shrinks the size and volume of the tetrahedral and octahedral interstitial sites in Ti-Mo alloys, which accordingly results in the reduction of the trapped hydrogen concentration. Additionally, a large amount of excess vacancies could be present in these nanostructured β phases, which can also serve as strong hydrogen trapping sites, forming vacancy-hydrogen clusters [30]. However, the addition of Mo would increase the formation energy of vacancy-hydrogen clusters, inhibiting hydrogen aggregation at vacancies [30]. This implies that the concentration of hydrogen trapping at vacancies may also decrease with increasing the Mo in the Ti-Mo alloys.

In summary, quantitative near-atomic-scale investigations of the distribution of H in a fully-β, as-homogenized and mixed α+β as-aged Ti-12Mo alloy were performed using APT complemented by electron microscopy. In the former sample, hydrogen either distributed homogeneously without notable hydrogen segregation to grain boundaries, or formed small agglomerations in the matrix, depending upon the total H content. For the latter sample, the hydrogen composition in β ranges from ~4% to ~13%, which is much higher than the approx. 2% in α. Hydrogen also segregates at some α/β interfaces to form hydrogen-enrichment layers with local concentrations exceeding 20%. Moreover, the hydrogen concentration roughly decreases linearly with increment of Mo content in β phases due to the suppression of hydrogen uptake by Mo alloying. No stable hydrides, such as

δ-, ϵ- and γ- titanium hydride, were found in either samples. The results help to further understand the behaviour of hydrogen in β and aged α+β titanium alloys.

Acknowledgements

Uwe Tezins & Andreas Sturm for their support to the FIB & APT facilities at MPIE. FKY is grateful for the National Natural Science Foundation (grants 51501191). AB and FKY are grateful for the funding received from the Alexander von Humboldt Foundation. LTS & BG are grateful for funding from the ERC-CoG SHINE – 771602. BG, LTS and IM acknowledge funding from the MPG through the Laplace project. YC is grateful to the China Scholarship Council (CSC) for funding of PhD scholarship